\documentclass[twocolumn,prb,showpacs,10pt,superscriptaddress]{revtex4-1}
\usepackage{textcomp}
\usepackage{amsmath}
\usepackage{amssymb}
\usepackage{subfig}

\usepackage{color}
\usepackage{multirow}
\usepackage{float} 
\usepackage{fontenc}
\usepackage{graphicx}
\usepackage{graphics}
\usepackage[justification=centering]{caption}
\usepackage{lipsum}
\makeatletter

\newcommand{\Rmnum}[1]{\expandafter\@slowromancap\romannumeral #1@}

\begin{document}

\title{Quantum interference in coherent tunnelling through branched molecular junctions containing ferrocene centers}
\author{Xin Zhao}
\affiliation{Institute for Theoretical Physics, TU Wien - Vienna University of Technology, Wiedner Hauptstrasse 8-10, A-1040 Vienna, Austria}
\author{Georg Kastlunger}
\affiliation{Institute for Theoretical Physics, TU Wien - Vienna University of Technology, Wiedner Hauptstrasse 8-10, A-1040 Vienna, Austria}
\affiliation{School of Engineering, Brown University, Providence, Rhode Island 02912, USA}
\author{Robert Stadler}
\email[Email:]{robert.stadler@tuwien.ac.at}
\affiliation{Institute for Theoretical Physics, TU Wien - Vienna University of Technology, Wiedner Hauptstrasse 8-10, A-1040 Vienna, Austria}

\date{\today}

\begin{abstract}
In our theoretical study where we combine a nonequilibrium Green's function (NEGF) approach with density functional theory (DFT) we investigate branched compounds containing ferrocene moieties in both branches which due to their metal centers are designed to allow for asymmetry induced by local charging. In these compounds the ferrocene moieties are connected to pyridyl anchor groups either directly or via acetylenic spacers in a meta-connection where we also compare our results with those obtained for the respective single-branched molecules with both meta- and para-connections between the metal center and the anchors. We find a destructive quantum interference (DQI) feature in the transmission function slightly below the lowest unoccupied molecular orbital (LUMO) which dominates the conductance even for the uncharged branched compound with spacer groups inserted. In an analysis based on mapping the structural characteristics of the range of molecules in our article onto tight-binding models, we identify the structural source of the DQI minimum as the through-space coupling between the pyridyl anchor groups. We also find that local charging on one of the branches only changes the conductance by about one order of magnitude which we explain in terms of the spatial distributions of the relevant molecular orbitals for the branched compounds.
\end{abstract}

\maketitle

\section{Introduction}\label{sec:intro}

Molecular electronics has become an active field of research in recent decades, since it holds the promise to maintain a continuous progress in the miniaturization of digital devices, thereby overcoming the limitations of semiconductor technology~\cite{ratner,loertscher}. One enabling tool for this purpose can be found in destructive quantum interference (DQI) effects~\cite{mayor,lambert1} which can significantly reduce the conductance in some conjugated $\pi$ systems, where this purely electronic effect has also been shown to be robustly observable at room temperature~\cite{molen}. For such hydrocarbon molecules a graphical atomic orbital (AO) scheme~\cite{graphical1}$^{-}$\cite{graphical5} as well as molecular orbital (MO) based rules~\cite{yoshizawa1}$^{-}$\cite{yoshizawa6} could be derived to predict the occurrence of DQI from the molecular structure, where the relation between the two schemes has been clarified recently~\cite{victor}. Such simplified schemes allow for the design of logical gates~\cite{graphical1} and memory cells~\cite{memory} in single molecule electronics as well as the implementation of thermoelectric devices~\cite{fano,lambert2}. 

Also the constructive quantum interference (CQI) expected in electron transport through branched molecular compounds gained attention, where a deviation from the classical Kirchhoff's law was first predicted theoretically~\cite{magoga} and then confirmed experimentally~\cite{joachim,vazquez} for junctions containing molecules providing symmetrically equivalent pathways through two of their branches. Recently, the design and synthesis of branched compounds containing ferrocene moieties in each branch has been presented~\cite{tim} for the purpose of creating single molecule junctions, where the combination of QI effects with redox gating for coherent electron tunneling as well as the electrostatic correlation between spatially distinct redox centers for electron hopping~\cite{hopping} can be explored. 

The latter electrostatic interactions between multiple ferrocene based redox centers within the same compound have been observed before in an unrelated study~\cite{tarraga}. Ferrocene moieties in junctions with linear molecules~\cite{kanthasamy} have also been used for the design of molecular diodes~\cite{engtrakul,nijhuits,ding}, highly conducting molecular wires~\cite{getty} and redox-gated molecular switches~\cite{xiao}, where the switching between a low-conductance reduced state and a high-conductance oxidized state was due to stochastic fluctuations between these two redox states induced by the gate. The details of the mechanism for this type of switching have recently been explored in joint experimental and theoretical studies on a Mo compound~\cite{hysterese1,hysterese2} and azulene~\cite{azulene}, where the I/V curves measured in a mechanically controlled break junction setup were also reproduced by simulations based on density functional theory (DFT).

\begin{figure}
\begin{center}
\includegraphics[width=1.0\linewidth,angle=0]{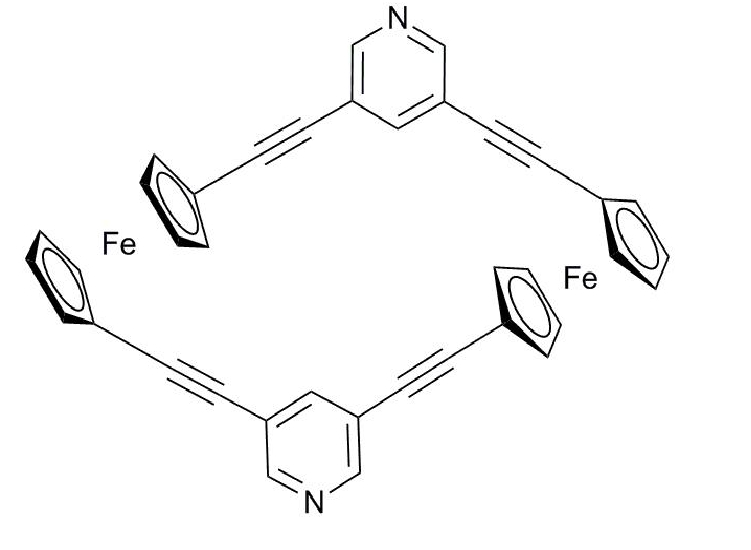}
\caption{\small Cyclic molecule containing a ferrocene moiety in each of its two branches, where they have been separated from the pyridyl anchor groups by acetylenic spacer groups.\label{fig1}}
\end{center}
\end{figure}

The novelty of the molecular design in Ref.~\cite{tim} lies in bringing all these structural aspects together in a single molecule, which could allow in principle to combine redox-gated fluctuations of the electron population at ferrocene moieties as a switching mechanism between two redox states, where DQI effects would guarantee a very low conductance for one of them and their absence a significantly higher conductance for the other one. A similar idea for a redox-gated switch, where one state of the redox pair was designed to exhibit DQI effects, has recently been pursued with anthraquinone derivatives but the ON/OFF ratios were found to be rather modest, since DQI occurred rather far in energy from the Fermi level (E$_F$) in the transmission function~\cite{anthraquinone}. Although the synthesis part of Ref.~\cite{tim} focused on branched compounds where both branches were to be attached on a substrate separately and only connected intra-molecularly by a pyridyl anchor group on the end to be contacted by the tip of a scanning tunneling microscope (STM), the authors stated in their conclusions that efforts towards cyclic analogues of these molecules such as the one shown in Fig.~\ref{fig1} were underway.

Such cyclic analogues are of particular interest in the context described above, since as pointed out in Ref.~\cite{joachim} QI effects can only be expected to play a dominant role for electron transport if both sides of a branched molecule are connected to electrodes by a common intra-molecular node. For the molecule in Fig.~\ref{fig1} in its neutral state the transmission through both branches is expected to interfere constructively, because the branches are symmetry equivalent~\cite{magoga,joachim,vazquez}. If one of the two ferrocene moieties is oxidized, however, this symmetry would be broken thereby possibly enabling a DQI induced suppression of the conductance. In that case the compound in Fig.~\ref{fig1} could be used as a molecular redox switch with very high ON/OFF ratios. In the molecular design the acetylenic spacers are meant to make the molecular structure more rigid and to increase the distance between the two electrodes for the prevention of through-vacuum tunneling and for separating the redox-active centers from the leads. The pyridyl anchor groups were chosen because they were found to provide the best junction formation and conductance properties in recent experimental~\cite{wong} and theoretical~\cite{pyridil1}$^{-}$\cite{pyridil4} studies.

\begin{figure}
\begin{center}
\includegraphics[width=\linewidth,angle=0]{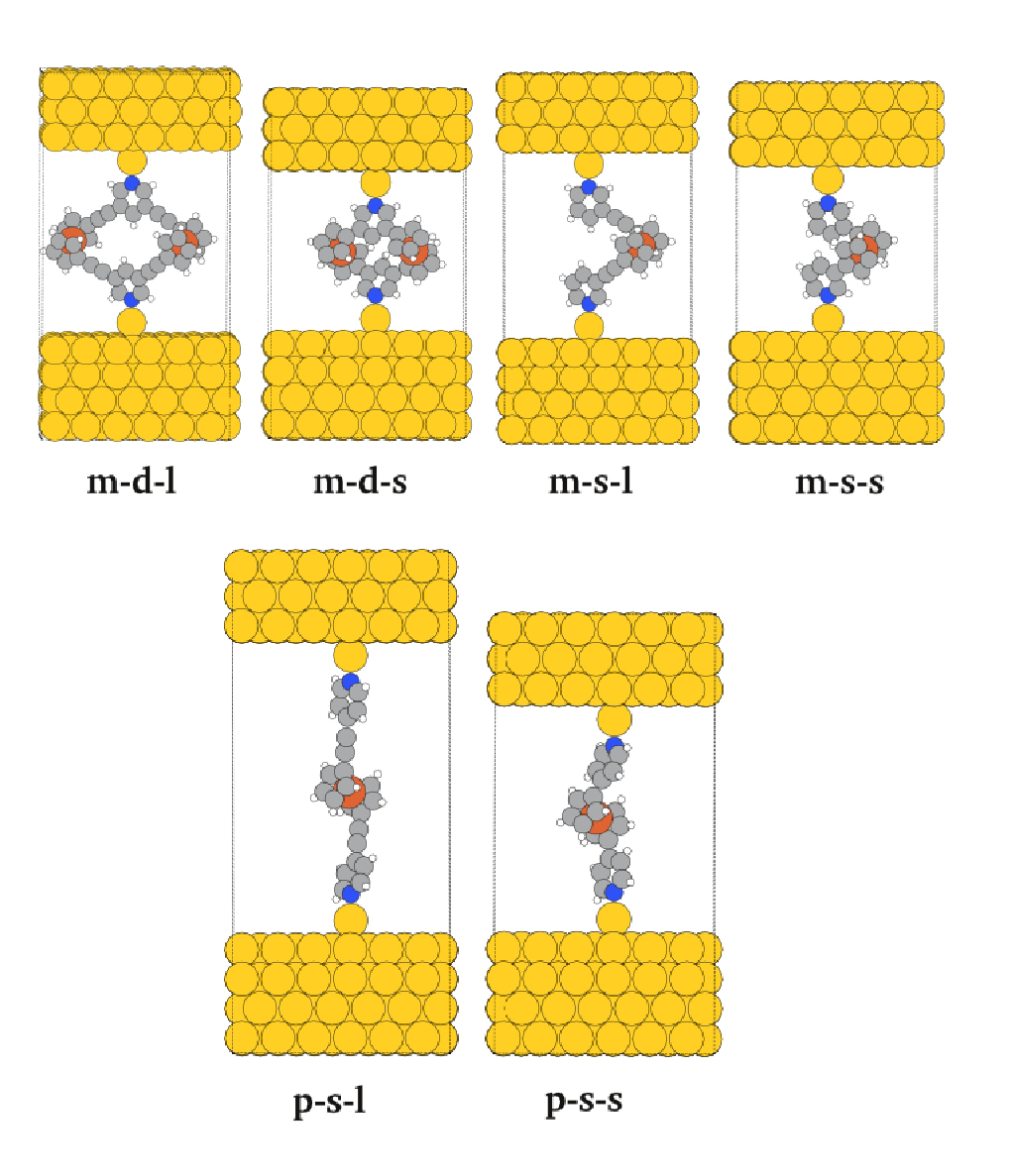}
\caption{\small Junction geometries for the compounds we investigate in this article, where m-d-l (meta-double-long) denotes the molecule in Fig.~\ref{fig1}, m-d-s (meta-double-short) the same molecule without the acetylenic spacer groups, m-s-l (meta-single-long) and m-s-s (meta-single-short) the corresponding single-branched compounds, and p-s-l (para-single-long) and p-s-s (para-single-short) their respective counterparts with a para-connection to the pyridyl anchor groups. All molecules have been connected to fcc Au electrodes on (111) surfaces with an ad-atom on each lead.\label{fig2}}
\end{center}
\end{figure}

In our article we investigate the coherent electron transport through the molecule in Fig.~\ref{fig1} by means of DFT calculations in combination with a non-equilibrium Green's function formalism (NEGF)~\cite{keldysh}, where we put an emphasis on DQI effects in the neutral and charged compound. Because of the presence of the ferrocene moieties in the compound neither the graphical AO scheme nor the MO rules mentioned above can be applied for this purpose, since both have been designed exclusively for the study of $\pi$ conjugated hydrocarbons~\cite{victor}, which is also true for the quantum circuit rules derived in Ref.~\cite{lambert3}. In the present case, however, DQI can arise i) from the meta-connection~\cite{meta1,meta2,meta3} of the branches to the pyridyl anchor group, although it has been recently demonstrated that for meta-connected bipyridine DQI in the $\pi$ electron contribution can be masked by the conductance mediated by $\sigma$ electrons~\cite{bipy}, ii) from interference between transmission through the two branches which is expected to be constructive for the neutral molecule but might be destructive if the redox-active center on only one of the branches is oxidized, and iii) also from multiple paths provided by nearly degenerate orbitals on the ferrocene moieties. In order to be able to distinguish between these effects we extend our study to the range of molecular junctions illustrated in Fig.~\ref{fig2} where derived from the compound in Fig.~\ref{fig1} we also chose molecules without acetylenic spacers, with only one branch between the pyridyl anchor groups and with para-connections for the single-branched systems.

The paper is organized as follows: In the next section we present transmission functions from NEGF-DFT~\cite{atk}$^{-}$\cite{kristian} calculations for all junctions in Fig.~\ref{fig2} and discuss their characteristic features. There we find that DQI occurs for neutral compounds in the energy region of the lowest unoccupied MO (LUMO) close to the Fermi level with a strong impact on the conductance only for molecules with branches connected in meta-positions at the pyridyl anchors with respect to their N atom and containing acetylenic spacers regardless of the number of branches, i.e. for the compounds we refer to as m-d-l and m-s-l in the caption of Fig.~\ref{fig2}. In Sec.~\ref{sec:sources} we derive topological tight-binding (TB) models from the DFT calculations and identify the through-space coupling between the pyridyl anchor groups which depends on both the meta- versus para-connectivity and the presence or absence of spacer groups as the defining quantity for the DQI effects we observe. In Sec.~\ref{sec:charging} we assess the usefulness of the double branched systems m-d-l and m-d-s in Fig.~\ref{fig2} as molecular switches by explicitly putting a positive charge on one of the two branches in our NEGF-DFT calculations and comparing the resulting conductance with that of the respective neutral compound. We conclude with a brief summary of our results in Sec.~\ref{sec:summary}.

\section{DFT based electron transport calculations and molecular orbitals for the neutral complexes}\label{sec:dft}

\begin{figure*}
    \begin{center} 
    \includegraphics[width=\linewidth,angle=0]{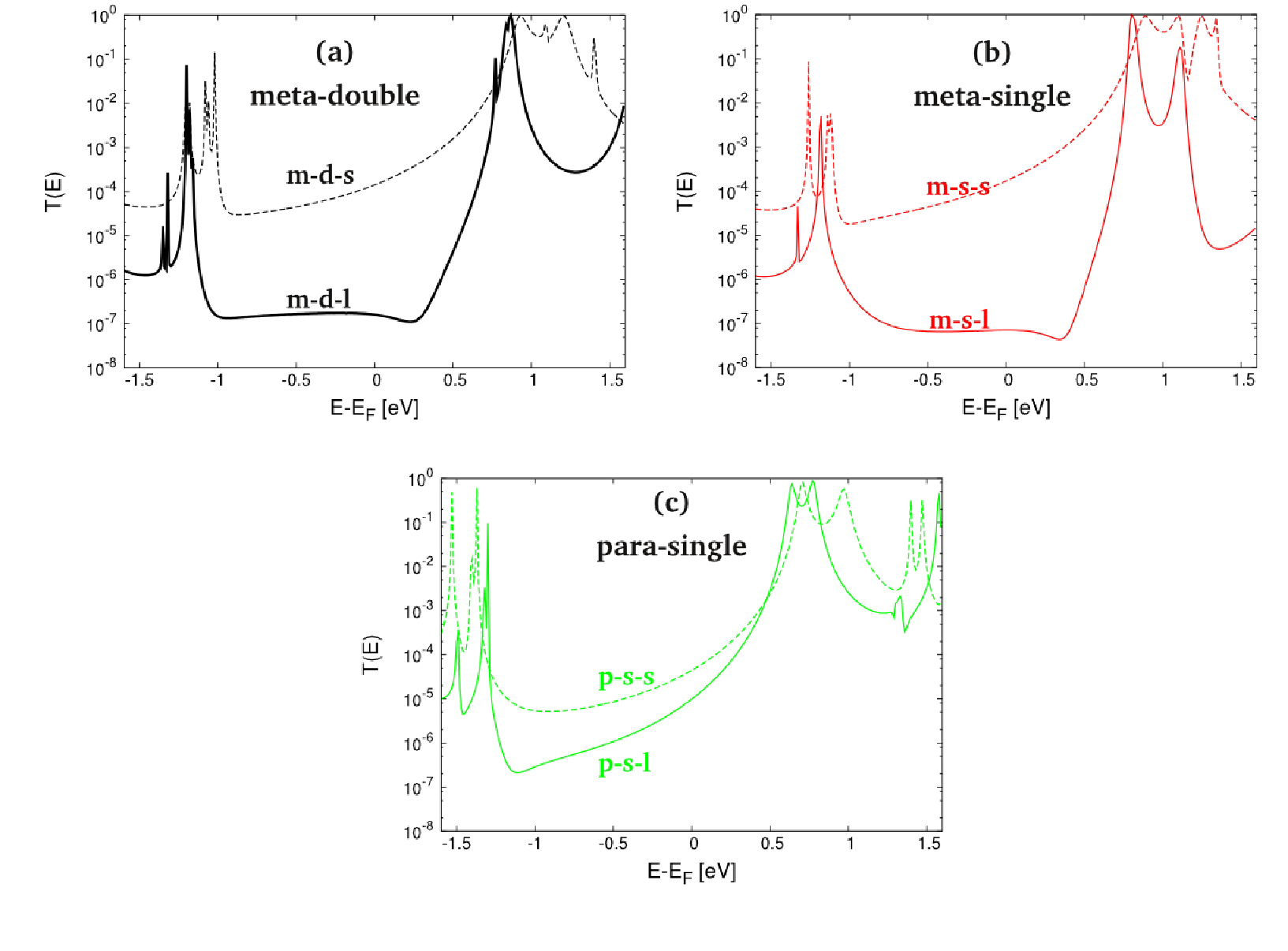}
    \caption{\small Transmission functions calculated from NEGF-DFT for the six junctions in Fig.~\ref{fig1}, where the color code for the lines is: a) m-d-l solid black, m-d-s dashed black, b) m-s-l solid red, m-s-s dashed red, c) p-s-l solid green and p-s-s dashed green.\label{fig3}}
    \end{center}
\end{figure*}

\subsection{Computational details for the NEGF-DFT calculations}

The transmission functions T(E) for all junctions in Fig.~\ref{fig2} we obtained from NEGF-DFT calculations performed with the GPAW code ~\cite{gpaw1,gpaw2} using a linear combination of atomic orbitals (LCAO)~\cite{lcao} for the basis set on a double zeta level with polarisation functions (DZP), a Perdew-Burke-Ernzerhof (PBE) parametrization for the exchange correlation (XC) functional~\cite{pbe} and a grid spacing of 0.2 \AA{} for the sampling of the potential in the Hamiltonian on a real space grid. In our transport calculations, the "extended molecule" defining the scattering region is formed by the respective metal organic compounds and three and four layers for the upper and lower fcc gold electrodes, respectively, in a (111) orientation and with a 6x6 over-structure defining the periodically repeated unit cell, where the distance between the Au ad-atom attached to the lead surfaces and the N atom of the pyridyl anchor groups was chosen as 2.12 \AA{}~\cite{pyridil1} and a $k$ points sampling corresponding to a 4x4x1 Monkhorst Pack grid for evaluating T(E), where the z-coordinate is the direction of electron transport through the junction. 

\subsection{Transmission functions from NEGF-DFT and the observation of DQI}

In the resulting transmission functions in Fig.~\ref{fig3} the peaks in the LUMO region are much broader than those in the HOMO region for all systems, and hence we expect the conductance to be dominated by the MOs above the Fermi level. DQI induced minima in the energy region at the upper border of the HOMO-LUMO gap can be observed only for meta-connected molecules with acetylenic spacers regardless of the number of branches but this feature disappears when the spacers are removed or when the connection of the ferrocene moieties to the pyridyl anchors is in a para-position. We note that these minima in T(E) in the LUMO region for the compounds m-d-l and m-s-l do not result in zero conductance accompanied by the typical DQI shape known from topological models~\cite{victor} but rather in a distinct deviation from a Lorentzian decay around the LUMO peaks which lowers the conductance significantly and has been encountered in molecules with meta-connected pyridyl anchors also in Ref.~\cite{lambert3}. 

Such less distinctly visible manifestations of DQI can occur in DFT calculations for real systems, because DQI is linked to the symmetry properties of the $\pi$-electrons of a conjugated system, where the $\sigma$-electrons are not necessarily affected~\cite{bipy}. Our definition of DQI is that the transmission through a system with more than one MO around E$_F$ is lower than the sum of the individual contributions of these MOs to T(E)~\cite{pyridil3}. The exact energetic position of the Fermi energy within the HOMO-LUMO gap, which is also affected by the underestimation of this gap in our calculations due to the PBE parametrization of the XC functional,  will have a crucial impact on the quantitative conductance but qualitatively DQI will always result in a significant conductance lowering for the structures where it occurs regardless of the details of the Fermi level alignment~\cite{victor}.

\subsection{General remark on CQI for the branched molecules}

One would expect from the circuit laws derived for branched molecules with two equivalent branches~\cite{magoga,joachim} that due to constructive QI the conductance of the molecules should be roughly four times as large as the respective value of the single branched analogue. While for the molecules containing acetylenic spacers we find indeed a ratio larger than two between the respective transmissions functions of m-d-l and m-s-l at E$_F$ in Fig.~\ref{fig3}, this is distinctly not the case for m-d-s and m-s-s where the conductance of the single-branched system is even slightly higher than the one found for the double-branched compound. In Refs.~\cite{lambert1} and~\cite{magoga} it was pointed out that the circuit laws for CQI only apply when the branches are rather weakly coupled to the nodal point in comparison with the nodal points electronic connection to the electrodes. In our case, however, the coupling between the ferrocene moieties and the pyridyl anchors is larger than the coupling between the anchors and the leads. In the experimental evaluation of the circuit laws for CQI in Ref.~\cite{vazquez} it was also found that the observability of these laws strongly depends on the chemical nature of both the anchors and the branches as well as on the atomistic details of the surface structure, the respective compounds are attached to.

\subsection{Molecular orbital analysis}

\begin{figure}
    \begin{center}
    \includegraphics[width=\linewidth]{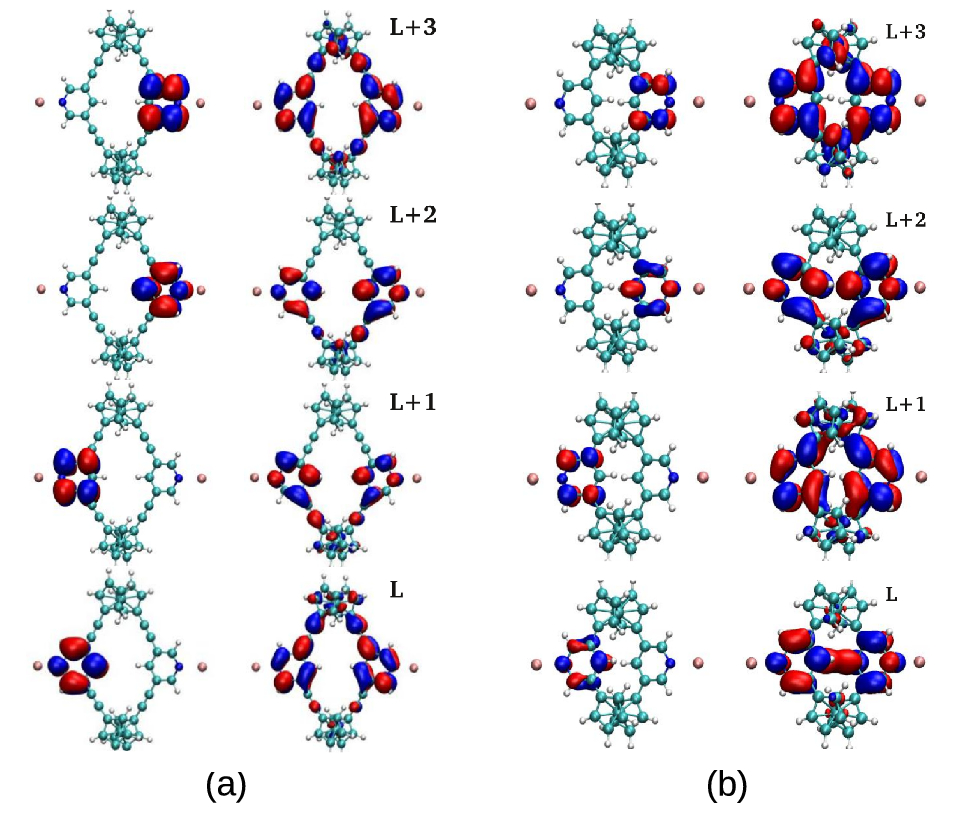}
    \caption{\small Spatial distributions of the four MOs directly above E$_F$ (LUMO,...,LUMO+3) for the branched compounds a) m-d-l and b) m-d-s, where the two FOs on each pyridyl anchor defining them are shown in the left panels and the four MOs themselves in the right panels.}
    \label{fig4}
    \end{center}
\end{figure}

\begin{figure}
    \begin{center}
    \includegraphics[width=\linewidth]{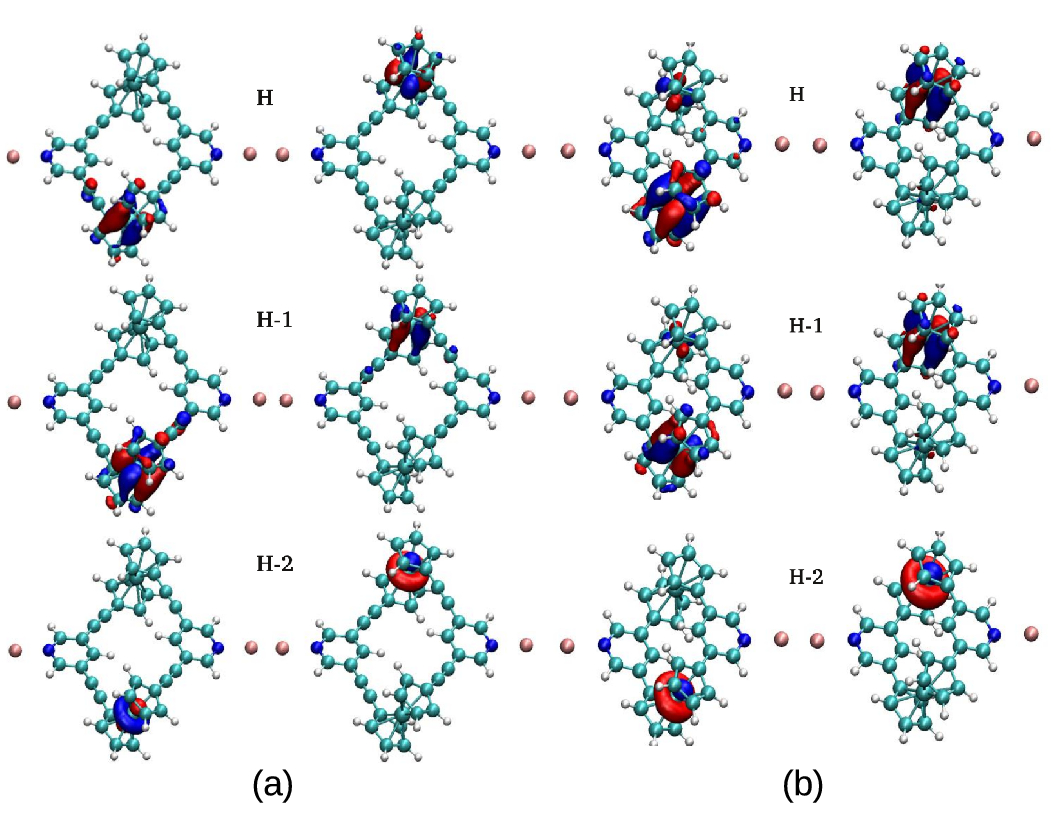}
    \caption{\small Spatial distributions of the six MOs directly below E$_F$ (HOMO,...,HOMO-5) for the branched compounds a) m-d-l and b) m-d-s, where the notation we use here (H,...,H-2) refers to each branch individually.}
    \label{fig5}
    \end{center}
\end{figure}

\begin{table}
\caption {Eigenenergies $\epsilon_{MO}$ in eV for the four MOs above E$_F$ (LUMO,...,LUMO+3) for all compounds in Fig.~\ref{fig3} and the three MOs below E$_F$ (HOMO,...,HOMO-2) for one of the two branches, where for m-d-l and m-d-s the values for the respective second branch are given in parentheses.} \label{tab:MOs}
\begin{center}
    \begin{tabular}{|c|c|c|c|c|c|c|}
      \hline
      & m-d-l  & m-d-s & m-s-l & m-s-s & p-s-l & p-s-s  \\
    \hline
    L+3  & 0.85  & 1.39   & 1.08   & 1.33 & 1.32 & 1.46  \\
    \hline
    L+2 & 0.81  & 1.17 & 1.06  & 1.21 & 1.28 & 1.39 \\
    \hline
    L+1 & 0.79 & 1.09 & 0.78  & 1.07 & 0.70 & 0.91\\
    \hline
    L & 0.76  & 0.86 & 0.76  & 0.81 & 0.57 & 0.64 \\
    \hline
    H & -1.16 (-1.20) & -1.02 (-1.06)  & -1.18  & -1.12 & -1.30 & -1.37 \\
    \hline
    H-1 & -1.18 (-1.21) & -1.02 (-1.08) & -1.19  & -1.14 & -1.32 & -1.41 \\
    \hline
    H-2 & -1.32 (-1.35) & -1.18 (-1.21) & -1.33  & -1.26 & -1.49 & -1.53 \\
    \hline
    \end{tabular}
\end{center}
\end{table}

In Figs.~\ref{fig4} and ~\ref{fig5} we plot the spatial distributions of the MOs for the double branched compounds directly above (Figs.~\ref{fig4}) and directly below (Figs.~\ref{fig5}) the Fermi energy, which we obtain from a subdiagonalization of the molecular part of the transport Hamiltonian~\cite{pyridil4}. In Table~\ref{tab:MOs} the corresponding eigenenergies are listed for all six junctions in Fig.~\ref{fig2}, where the lists are complete for the energy range -2 eV $<$ E$_F$ $<$ 1.5 eV and the shapes of MOs for the single-branched systems share the same localization patterns with those plotted for the double branched molecules in Figs.~\ref{fig4} and ~\ref{fig5}. For the LUMO region all four MOs are mostly localized on the pyridyl anchor groups (Fig.~\ref{fig4}), which explains the broad peaks found for all junctions in T(E) above E$_F$ (Fig.~\ref{fig3}). Furthermore, visual inspection allows to identify these four MOs as bonding/antibonding pairs resulting from the hybridization of just two pyridyl fragment orbitals (FOs), which were again obtained from a subdiagonalization of the respective transport Hamiltonian but in this case limited to the basis functions centered on the pyridyl groups.

The MOs below E$_F$ on the other side (Fig.~\ref{fig5}) are all mostly localized on the ferrocene moieties as hybrids of Fe d-states and the $\pi$-system of their cyclopentadienyl rings. As a result we observe rather narrow peaks of T(E) in the HOMO region (Fig.~\ref{fig3}) for all junctions. For the double-branched compounds, these six MOs can be clearly separated into three on each branch, suggesting that our initial concept of introducing a positive charge on one ferrocene center for the creation of an asymmetry resulting in DQI might work for T(E) in the HOMO region. It is, however, not likely to be applicable in the LUMO region since the localization patterns on the pyridyl anchors cannot be expected to be affected in an asymmetric way by the charging of one ferrocene moiety. In order to evaluate the validity of this first assessment further, NEGF-DFT calculations with explicitly charged ferrocene centers will be presented in Sec.~\ref{sec:charging}.

Since both the shapes and eigenenergies (as listed in Table~\ref{tab:MOs}) of all MOs are quite similar for the six junctions in Fig.~\ref{fig2}, it cannot be directly derived from these properties why a DQI feature occurs in T(E) in the LUMO region for compounds m-s-l and d-s-l but not for the other four molecules in Fig.~\ref{fig3}. It might be expected that the number of branches does not make a difference here because the existence of the second branch should induce CQI but not DQI without the charging of a ferrocene center~\cite{magoga,joachim,vazquez}. It also seems intuitive that molecules, where ferrocene is connected to the pyridyl anchors in meta-connections exhibits DQI while the para-analogue does not but this intuition is only based on the observations made for planar $\pi$ conjugated hydrocarbons~\cite{meta1,meta2}, while the six compounds in Fig.~\ref{fig2} are not planar and contain ferrocene moieties. Most strikingly, there is no easy explanation for the dependence of the DQI feature on the absence or presence of acetylenic spacers. In order to investigate these questions further, we project the data we can derive from NEGF-DFT calculations onto topological TB models in Sec.~\ref{sec:sources}.

\section{Investigation of the structural sources of DQI with TB models}\label{sec:sources}

All conventional topological TB models and the various sets of QI or quantum circuit rules derived from such models have been developed for planar $\pi$-conjugated hydrocarbons. Also the simple assertion that meta-connected junctions exhibit DQI, while para-connected ones do not, can be considered to be a simple case of a QI rule derived from a conventional topological TB model. In such models the molecular structure is replaced by a connectivity matrix where each carbon position is represented by a single AO (presumably the p$_z$ orbital perpendicular to the plane), where all AOs have the same onsite energy and only next neighbor couplings are considered. The Ferrocene makes both assumptions ambiguous. It cannot be represented by carbon p$_z$ AOs alone but has degenerate FOs at different energies instead. Additionally, it enforces molecular structures in three dimensions in deviation from planarity, where parts of the molecule not directly bonded to each other can come close to each other in the third dimension and QI cannot be understood in terms of next-neighbor connectivity alone anymore. With conventional TB models being not applicable for the structures we investigate, we have to derive our own models, which have to fulfill two requirements: i) the qualitative structure dependence of the transmission functions from our DFT calculations needs to be reproduced, and ii) the number of orbitals involved at the end needs to be minimal in order to make the key structural source of DQI in our systems discernible. Such a step-by-step model development is introduced in this section and the applicability of this procedure is not just narrowly limited to the particular six molecules we investigate but is also given for similar systems.

For the double-branched molecules the transmission functions in Fig.~\ref{fig3} have a very similar shape to that of their respective single-branched analogues in the LUMO region in the meta-connected cases and the acetylenic spacers do not seem to have a significant impact for para-connected anchors other than the well-known decrease of the conductance with molecular length. Therefore, we focus our analysis of the relationship between structural features and T(E) in this section on an evaluation of the differences between compounds m-s-l, m-s-s and p-s-l. 

\subsection{Definition of the electrodes for all NEGF-TB calculations}

We calculate transmissions functions from NEGF-TB with a one-dimensional chain of AOs acting as electrodes where all inner electrode onsite energies have been set to 0.83 eV and all inner electrode couplings to -5.67 eV. This particular choice for the latter two parameters has been identified as optimal for reproducing NEGF-DFT results for T(E) with fcc Au (111) electrodes in Ref.~\cite{pyridil3} and will be used for all NEGF-TB calculations in our current article. The couplings between the contact atoms of these artificial electrodes with the p$_z$ orbitals within the pyridyl anchors have been derived by subdiagonalizing part of the transport Hamiltonian from the NEGF-DFT calculations describing the gold ad-atoms on top of the surfaces (see Fig.~\ref{fig2}) and taking only the couplings of the valence s state of this atom to the pyridyl p$_z$ states because the density of states of the gold surface has predominant s-character around E$_F$.

\subsection{Selection of AOs on the anchor groups and FOs on the ferrocene for reproducing the DFT results}

In the first part of this analysis we try to map the structural characteristics of these three molecules onto a topological TB model with the aim to match T(E) from NEGF-DFT as closely as possible but at the same time minimize the number of involved orbitals. 

For the pyridyl anchors and the acetylenic spacers it can be safely assumed that transport near the HOMO-LUMO gap is dominated by the p$_z$ AOs on the C and N sites~\cite{pyridil3}. The DZP-LCAO basis set of the NEGF-DFT calculations, however, does not provide physically meaningful AOs on particular atoms in the environment of all neighbouring atoms. Hence, we obtained the basis which we need to apply in our TB models by subsequent subdiagonalizations and basis set rotations of the transport Hamiltonian on each C and N atom individually.~\cite{bipy} Additionally, orthogonality between AOs on neighbouring atoms was ensured by applying a L\"{o}wdin transformation~\cite{loewdin}. As it has been demonstrated in the supporting information of Ref.~\cite{bipy} that not only first but also second and third nearest neighbor couplings within a pyridyl group are crucial for defining the energetic position of a DQI minimum, we include all three categories in our model. For the ferrocene part of the molecules we perform a subdiagonalization of the part of the Hamiltonian covering the whole moiety which results in just five FOs in the relevant energy range around the HOMO-LUMO gap, namely three FOs with energies from -1.55 to -1.20 eV in the HOMO region and two FOs with energies from 1.35 to 1.55 eV in the LUMO region for all three compounds.

\begin{figure}
    \begin{center}
    \includegraphics[width=\linewidth]{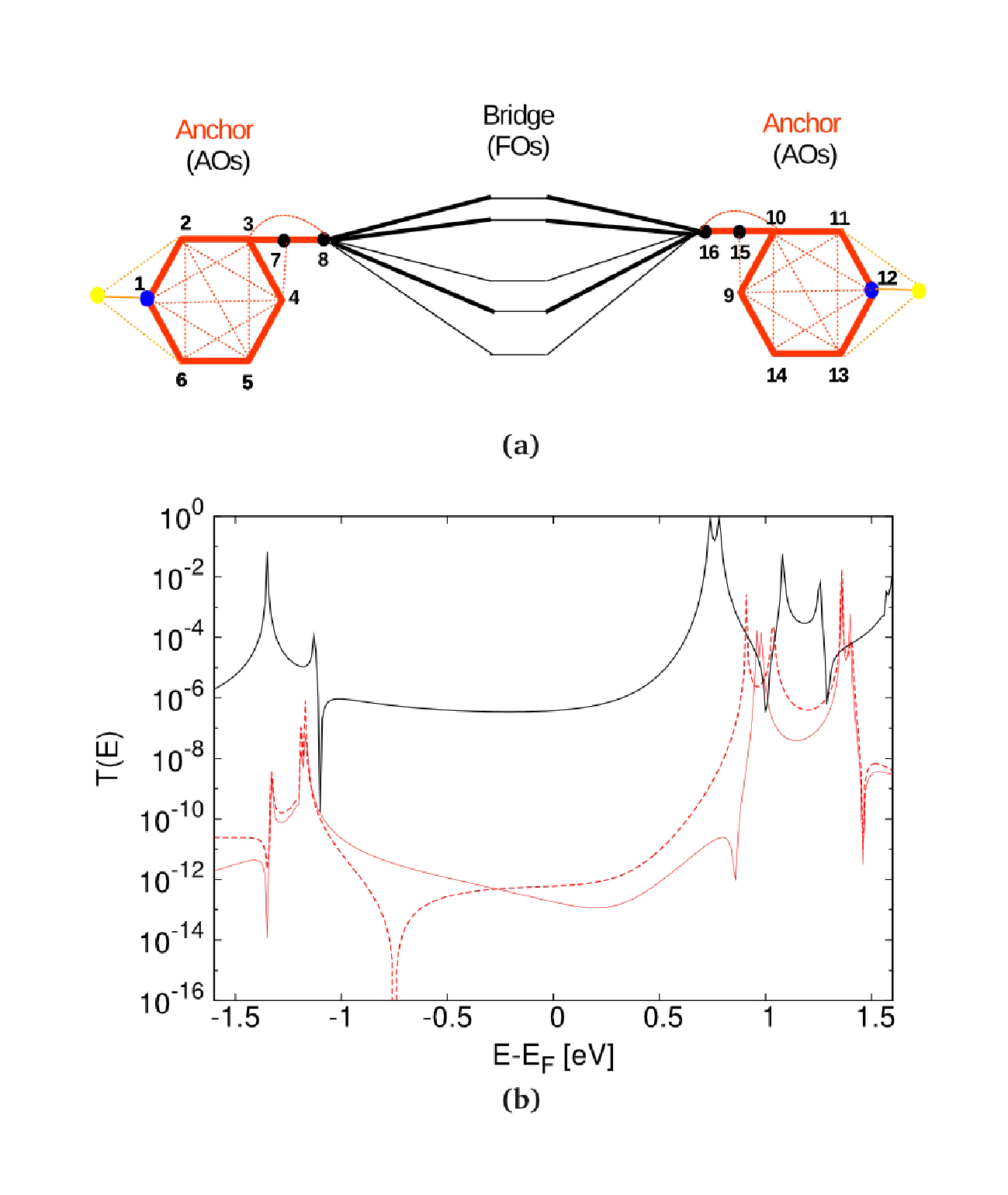}
    \caption{\small a) TB model derived from NEGF-DFT as exemplified for compound m-s-l, where the line colors distinguish between couplings within anchors and spacers (red) and couplings to ferrocene FOs (black) or the gold leads (yellow), respectively, while the line thickness tries to illustrate a hierarchy of the respective coupling strengths, b) T(E) from the TB model for $m-s-l$ (red) with (dashed curve) and without (solid curve) an artificially high value for the direct coupling and for $m-s-s$ (black solid curve).}
    \label{fig6}
    \end{center}
\end{figure}

In Fig.~\ref{fig6}a we illustrate this TB model schematically for molecule m-s-l where direct couplings between the p$_z$ orbitals of anchor and spacer groups left and right of the ferrocene are not drawn for the sake of simplicity but still considered in the model.

\subsection{Identification of the through-space coupling between anchor groups as the structural source of DQI}

Since we know that the most distinct structural difference between molecules m-s-l and m-s-s lies in their respective molecular length as brought about by the presence or absence of the acetylenic spacers, we show in Fig.~\ref{fig6}b T(E) from NEGF-TB for m-s-l in the original parametrization as derived from DFT (solid red line) and with just one parameter changed to the higher value we obtain for m-s-s (dashed red line), namely the direct coupling between the AOs 4 and 9 in Fig.~\ref{fig6}a. Of course, this "artificial" parametrization, which is meant to mimic a key structural aspect of m-s-s does not reproduce the high conductance found for this system in Fig.~\ref{fig3} but it can be clearly seen that just changing this one parameter from the value it has in m-s-l to the one it has in m-s-s seems to be sufficient to shift the DQI feature so far down in energy that it is not observable in the LUMO region anymore. In Fig.~\ref{fig6}b we also plot the transmission function we obtain from the parameters and topology of compound m-s-s (black solid line) which just like the one for m-s-l (red solid line) perfectly reproduces all characteristics found from NEGF-DFT in Fig.~\ref{fig3}. The model, however, needs to be simplified further in order to pin down and separate the effects of the most important structural differences between the single-branched molecules.

\begin{figure*}
    \begin{center}
    \includegraphics[width=\linewidth]{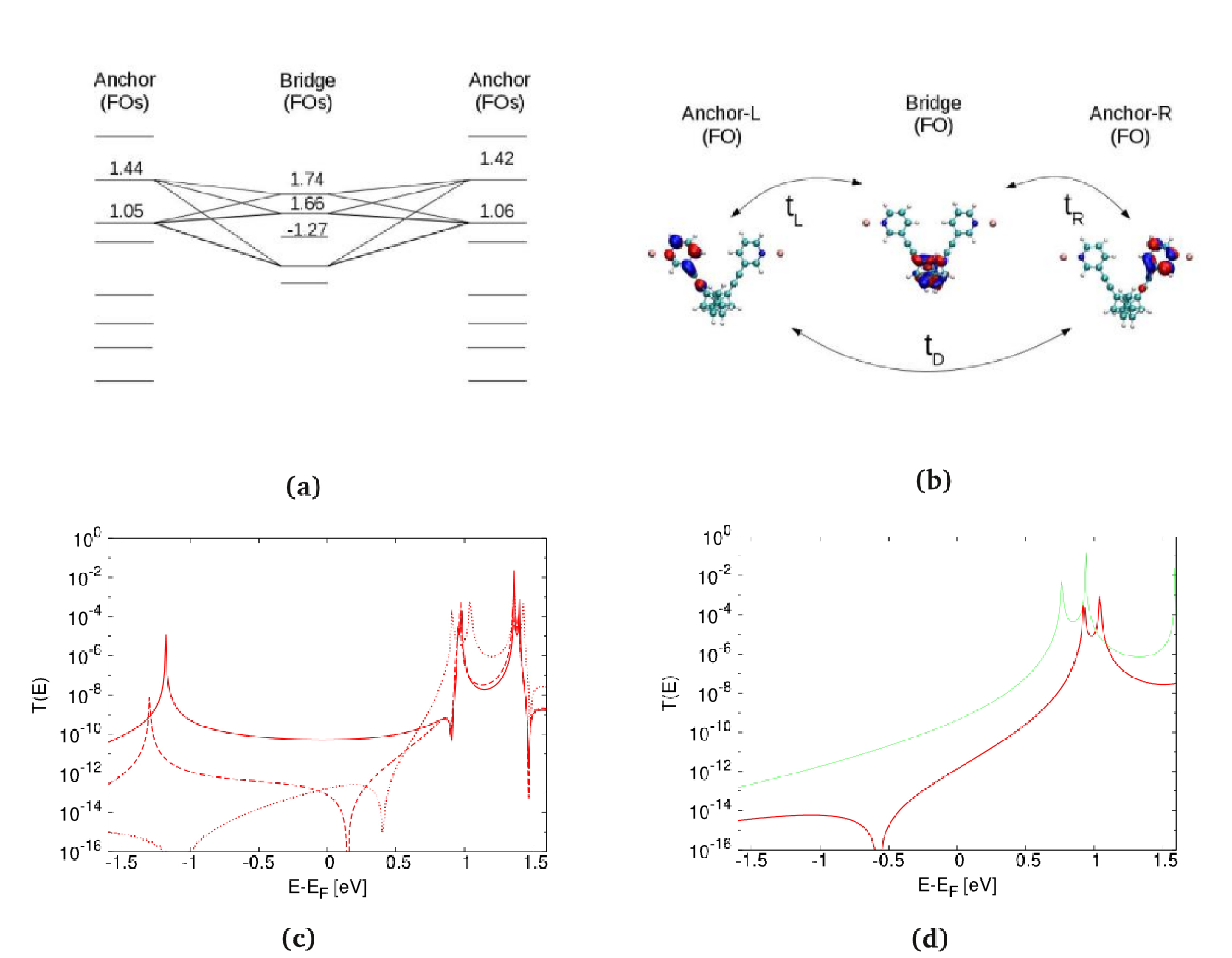}
    \caption{\small (a) FO-TB model for molecule $m-s-l$ as described in the main text, where the relevant couplings between anchor and bridge states which are all in the range of 0.1-0.2 eV have been indicated with lines and the spatial distributions of the anchor FO on each side at 1.05 eV and the ferrocene FO at 1.66 eV are shown in b). T(E) as calculated from NEGF-TB for this model are shown as red lines in c) for two anchor FOs (as marked in a)) and all five bridge FOs (solid), only three bridge FOs (as marked in a), dashed) and only the one bridge FO at 1.66 eV (dotted), and in d) for compound $m-s-l$ (red) and $p-s-l$ (green) for only the one anchor FO on each side and one bridge FO in the middle as plotted in b).}
    \label{fig7}
    \end{center}
\end{figure*}

For that purpose we perform another subdiagonalization of the transport Hamiltonian in the subspace of the eight p$_z$ AOs on the pyridyl anchors and acetylenic spacers in Fig.~\ref{fig6}a on each side of the ferrocene center. This results in the FO-TB model in Fig.~\ref{fig7}a, where the five FOs on the ferrocene moiety are the same as in Fig.~\ref{fig6}a and two FOs on each anchor can be roughly identified from their shape with those shown in Fig.~\ref{fig4}a, albeit they now show some localization on the spacer groups too due to the way of their definition. From the size of the couplings of the five bridge FOs to these two anchor FOs we can identify the three bridge FOs most relevant for the molecule m-s-l, namely one in the HOMO region and two in the LUMO region as indicated in Fig.~\ref{fig7}a while for p-s-l only the bridge FO lowest in energy in the LUMO region and only the lower lying of the two FOs on the anchors plays a role for the transmission. 

Molecules m-s-l and p-s-l now differ in the FO-TB model in two ways, namely in the number of FOs on each of the three fragments connected by sizable couplings and in the detailed values for these couplings. Therefore, the question arises if DQI in T(E) would still be found for compound m-s-l if only the one FO on each fragment also relevant for system p-s-l but with the parameters for m-s-l (Fig.~\ref{fig7}b) is selected for NEGF-TB calculations with a minimal number of FOs. In Figs.~\ref{fig7}c and d, we present the results of such calculations where Fig.~\ref{fig7}c shows T(E) for molecule m-s-l with two anchor FOs on each side and five (solid line), three (dashed line) and one FO (dotted line) on the ferrocene, respectively, and it can be seen that the DQI feature is shifted to the HOMO region if the quality of the FO model is reduced but remains observable. In Fig.~\ref{fig7}d we choose the same one FO on each fragment setup for compounds m-s-l (red curve) and p-s-l (green curve) as illustrated in Fig.~\ref{fig7}b, where we come to the somewhat surprising conclusion that still DQI is observed for m-s-l but not for p-s-l although the models for the two systems now only differ in the detailed parameters for the couplings between three FOs which have very similar spatial distributions and onsite energies for both cases.

\subsection{Analysis of the mathematical reasons for the decisive influence of the through-space coupling with a simplified 3x3 Hamiltonian}

\begin{figure}
    \begin{center}
    \includegraphics[width=1.0\linewidth,angle=0]{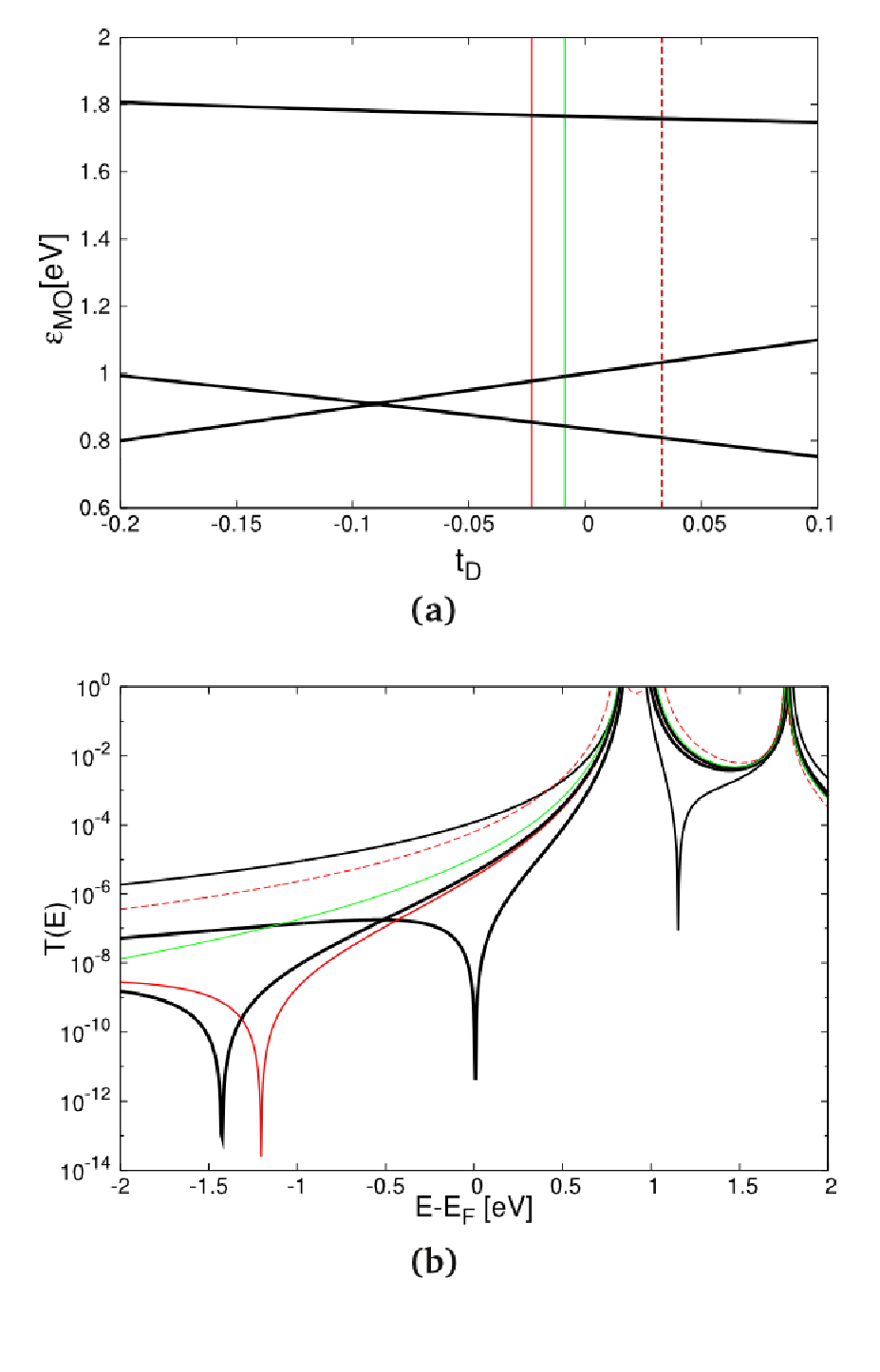}
    \caption{\small (a) MO Eigenenergies obtained by diagonalizing the 3x3 Hamiltonian formed by the three FOs in Fig.~\ref{fig7}b with $\epsilon_L$=$\epsilon_R$=1.0 eV, $\epsilon_B$=1.6 eV, t$_L$=0.25 eV, t$_R$=-0.25 eV and t$_D$ as a variable, where the three vertical lines mark the respective $t_{D}$ values of compounds $m-s-l$ (red solid line), $p-s-l$ (green solid line) and $m-s-s$ (red dashed line), (b) transmission functions calculated from $\Gamma^2 (E)$ for the resulting three MOs for the $t_{D}$ values of the three molecules explicitly given in Table~\ref{tab:couplings} and highlighted in the colors corresponding to a) and as black lines for t$_D$=-0.14, -0.04 and -0.02 eV, respectively.} 
    \label{fig8}
    \end{center}
\end{figure}

\begin {table}
\setlength{\tabcolsep}{1.1em}
\caption {Couplings connecting the three FOs in Fig.~\ref{fig7}b for three of the single branched systems, where all values are given in eV.}\label{tab:couplings}
\begin{center}
    \begin{tabular}{|c|c|c|c|}
    \hline
      Couplings & $m-s-l$ & $p-s-l$ & $m-s-s$   \\
    \hline
     $t_L$ & 0.27  & -0.23   & -0.28    \\
    \hline
     $t_R$ & -0.22  & 0.25 & 0.22   \\
    \hline
    $t_D$ & -0.023 & -0.0087  & 0.033 \\
    \hline
    \end{tabular}
\end{center}
\end {table}

In Table~\ref{tab:couplings} we list the coupling values for t$_L$, t$_R$ and t$_D$ connecting the three FOs in Fig.~\ref{fig7}b for all three junctions, where the first two parameters do not vary with the molecular structure significantly but the third one does. Having now established that the direct coupling between the two anchor groups distinguishes the only single-branched system with a DQI feature close to the LUMO, namely molecule m-s-l, from both compounds m-s-s and p-s-l, we want to explore the mathematical reasons for the importance of this parameter. We therefore diagonalized a 3x3 Hamiltonian with fixed parameters for the three FOs in Fig.~\ref{fig7}b and plotted the evolution of the resulting three MOs in dependence on t$_D$ in Fig.~\ref{fig8}a. In Fig.~\ref{fig8}b we show the transmission functions for selected values of t$_D$, which we obtained by making use of Larsson's formula~\cite{lars}
\begin{equation}\label{larsson}
 \Gamma (E)= \sum_{i} \frac{\alpha_{i}\cdot \beta_{i}}{E-\varepsilon_{i}}
\end{equation}
where $\varepsilon_{i}$ is the eigenenergy of each MO, and $\alpha_{i}$ and $\beta_{i}$ its respective coupling to the left and right electrode. Larsson's formula was originally introduced for the definition of the transfer integral in the context of Marcus theory for the description of electron hopping \cite{lars,lars1,lars2}, but recently it has been shown that it can be also used for approximating $T(E)$ as $T(E)$ $\sim$ $\Gamma^2 (E)$ for coherent tunneling \cite{lars3,pyridil3}, where the resulting T(E) can be normalized~\cite{lars3} and qualitatively reproduces the curves obtained from NEGF-TB~\cite{victor}. 

Equation~\ref{larsson} has the additional advantage that a simple mathematical condition can be defined for the energetic positions of DQI induced zeros in T(E), because at the same energies the effective coupling  $\Gamma (E)= \gamma_{1}/(E-\varepsilon_{1})+\gamma_{2}/(E-\varepsilon_{2})+\gamma_{3}/(E-\varepsilon_{3})$ with $\gamma_i=\alpha_i \beta_i$ for the three MOs resulting from the simple model in Fig.~\ref{fig8} must also be zero. By making use of the specific symmetry properties of the 3x3 Hamiltonian in the model, we can impose $\gamma_1+\gamma_2+\gamma_3$=0 and obtain 
\begin{equation}\label{E0}
 E_{0}= \varepsilon_{1} + \frac{1}{1+\frac{\gamma_{3} (\varepsilon_{3}-\varepsilon_{2})} {\gamma_{1} (\varepsilon_{1}-\varepsilon_{2})}}\textperiodcentered{(\varepsilon_{3}-\varepsilon_{1})} = \varepsilon_{1} + F_1 F_2
\end{equation}
for the energy of the DQI induced minimum, i.e. the energy E$_0$ defined by the condition T(E$_{0}$)=0 in our model.

\begin {table}
\setlength{\tabcolsep}{1.1em}
\caption {Explicit values for all parameters entering Equation~\ref{E0} for the three MOs obtained by diagonalizing the 3x3 Hamiltonian formed by the three FOs in Fig.~\ref{fig7}b with $\epsilon_L$=$\epsilon_R$=1.0 eV, $\epsilon_B$=1.6 eV, t$_L$=0.25 eV, t$_R$=-0.25 eV and t$_D$ as a variable. All values for t$_D$ and E$_0$ are given in eV, while the factors are dimensionless.}\label{tab:factors}
\begin{center}
    \begin{tabular}{|c|c|c|c|}
    \hline
       & $m-s-l$ & $p-s-l$ & $m-s-s$   \\
    \hline
     t$_D$ & -0.023 & -0.0087 & 0.033 \\
    \hline
     $E_{0}$ & -1.12 & -5.58  & 3.49 \\
    \hline
     $\gamma_{3}/\gamma_{1}$ & 0.225  & 0.218 &  0.20 \\
    \hline
     $F_{splitting}$ & -6.49  & -5.24 & -3.24 \\
    \hline
     $F_1$ & -2.16 &  -6.97  &  2.83 \\
    \hline
     $F_2$ & 0.91  & 0.92 &  0.95 \\
    \hline
    \end{tabular}
\end{center}
\end {table}

\begin{figure}
    \begin{center}
    \includegraphics[width=1.0\linewidth,angle=0]{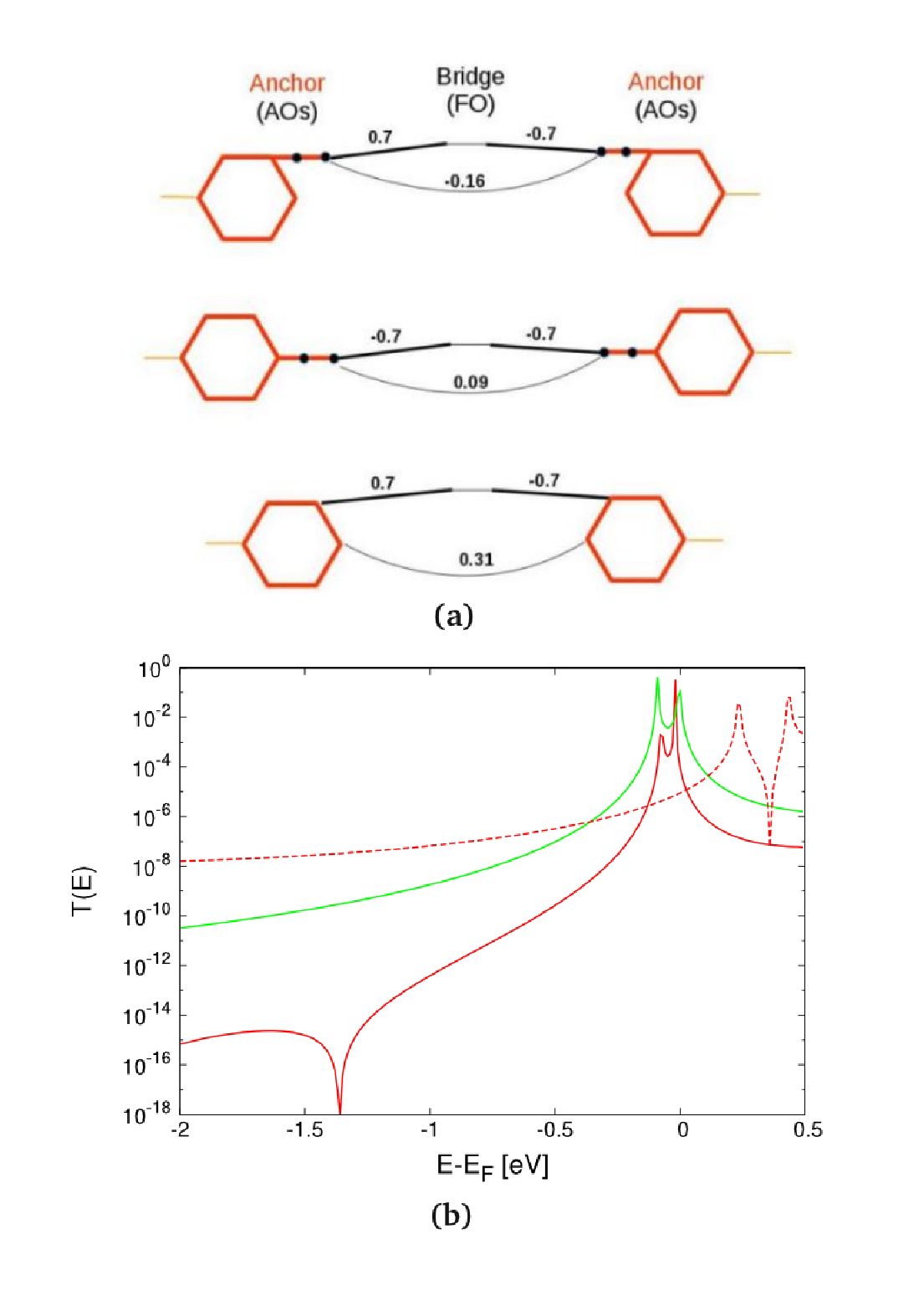}
    \caption{\small a) Conventional topological TB-models for compounds $m-s-l$, $p-s-l$ and $m-s-s$ where only next neighbor couplings have been considered within the pyridyl groups and acetylenic spacers which are all set to -3.0 eV while the coupling of the contact atom to the leads is taken to be -0.2 eV. For the onsite energies all AOs representing C and N sites are at -2.5 eV, while the single ferrocene FO is positioned at 1.7 eV for all three molecules, where only the sign of the couplings of this FO to the AOs nearest to it differs between meta and para and the value for the direct through-space coupling between anchor FOs (both given explicitly in the figure) is different for all three structures. b) NEGF-TB calculations for the models in a) for $m-s-l$ (solid red), $p-s-l$ (solid green) and $m-s-s$ (dashed red).}
    \label{fig9}
    \end{center}
\end{figure}

In Equation~\ref{E0} the factor $F_2=\varepsilon_{3}-\varepsilon_{1}$ is always positive by definition since the indices order the MOs in the sequence of their respective eigenenergies. Therefore, it is the sign of the other factor in the product, namely $F_1=1/(1+{(\gamma_{3}/\gamma_{1})}\cdot((\varepsilon_{3}-\varepsilon_{2})/(\varepsilon_{1}-\varepsilon_{2})))$ which decides whether the minimum $E_0$ lies to the left or to the right of the LUMO's energy $\varepsilon_1$ on the energy axis. All three compounds discussed in this section have t$_D$ values to the right of the crossing point between the lower two MO energies in Fig.~\ref{fig8}a, i.e. higher than t$_D$=-0.09 eV, which we obtain from Table~\ref{tab:couplings} and list again in Table~\ref{tab:factors}. Within this range of t$_D$ ${\gamma_{3}}/{\gamma_{1}}$ is always positive and $F_{splitting}=(\varepsilon_{3}-\varepsilon_{2})/(\varepsilon_{1}-\varepsilon_{2})$ always negative, and therefore the product of these two latter factors must be always negative. Hence, the sign of $F_1$ is determined by whether this product is larger or smaller than 1, where we can see from Table~\ref{tab:factors} that ${\gamma_{3}}/{\gamma_{1}}$ is fairly system independent while $F_{splitting}$ varies widely. 

For molecules m-s-l and p-s-l where $E_0$ as a consequence of a negative $F_1$ lies to the left of the LUMO peak, the size of $F_{splitting}$ also determines how close in energy $E_0$ and this peak are, since $F_1$ scales inversely with $F_{splitting}$. The dependence of $F_{splitting}$ on t$_D$ can be directly read from Fig.~\ref{fig8}a where it can be seen that $F_{splitting}$ increases when the crossing point at -0.09 eV is approached from either side of the t$_D$ axis. We further illustrate this point in Fig.~\ref{fig8}b, where we plot $\Gamma^2 (E)$ in dependence on t$_D$ and find that T(E) is reproduced for the particular values for the three single-branched molecules. In addition we also pick two characteristic values to the right of the crossing point, where it can be seen that the one approaching it closer at -0.04 eV results in a DQI closer to the LUMO peak than the one further away at -0.02 eV or the value for compound m-s-l (-0.023 eV). With the t$_D$ value left from the crossing point at -0.14 eV we demonstrate that in this range $\gamma_3/\gamma_1$ becomes negative which means that $F_1$ is always positive thereby moving the DQI feature to energies higher than the LUMO peak, while $F_{splitting}$ then merely determines the energetic distance between the minimum and the peak.

\subsection{Introducing the the through-space coupling as an ad-hoc parameter into conventional topological TB models}

Now armed with the knowledge that the direct coupling $t_D$ for the FO model in Fig.~\ref{fig7}b reflects the structural differences most relevant for the occurrence or absence of the DQI feature below the LUMO peak for the range of molecules we investigate in this article, we return to the topological TB model we started from in Fig.~\ref{fig6}a and simplify it accordingly by removing all second and third nearest neighbor couplings within the anchor groups and all but one of the ferrocene FOs. In the resulting minimal topological TB-model (Fig.~\ref{fig9}a) we put all C and N sites at the same onsite energies for all compounds as well as using the same value for the next nearest neighbor couplings within all anchor groups. The single remaining ferrocene FO has an onsite energy higher than those of the AOs but also here the same value is chosen for all three systems. They now differ only in the direct coupling between the AOs on the anchor groups on opposite sides of the ferrocene closest to each other, and meta and para are also distinct in the signs of the couplings of these AOs to the bridge FO. These minimal structural differences in the model already fully reproduce the characteristic features in T(E) for all molecules as can be verified from the NEGF-TB calculations presented in Fig.~\ref{fig9}b.

\subsection{Conclusions from the TB analysis}

In summarizing this section, it can be said that molecules containing ferrocene moieties differ distinctly from planar conjugated hydrocarbons in the correspondence between molecular structure and DQI effects in electron transmission, where general rules derived from simplified topological assumptions for the latter~\cite{victor,lambert3} are not applicable for the former. Strikingly, the most important structural difference of the molecules in this study is not defined by either the meta- or para-connection of their respective components, the availability of almost degenerate orbitals on the ferrocene or the number of branches connecting the two anchor groups, although all of these aspects play a certain role in the exact energetic positioning of the DQI minimum. It is rather the direct through-space coupling between the anchor groups defined by the three-dimensional conformation of the respective compound and widely adjustable by spacer groups which determines the observability of DQI in T(E) in a delicate way.

\section{Effect of charging of the branched compounds}\label{sec:charging}

\begin{figure*}
    \begin{center}
    \includegraphics[width=1.0\linewidth,angle=0]{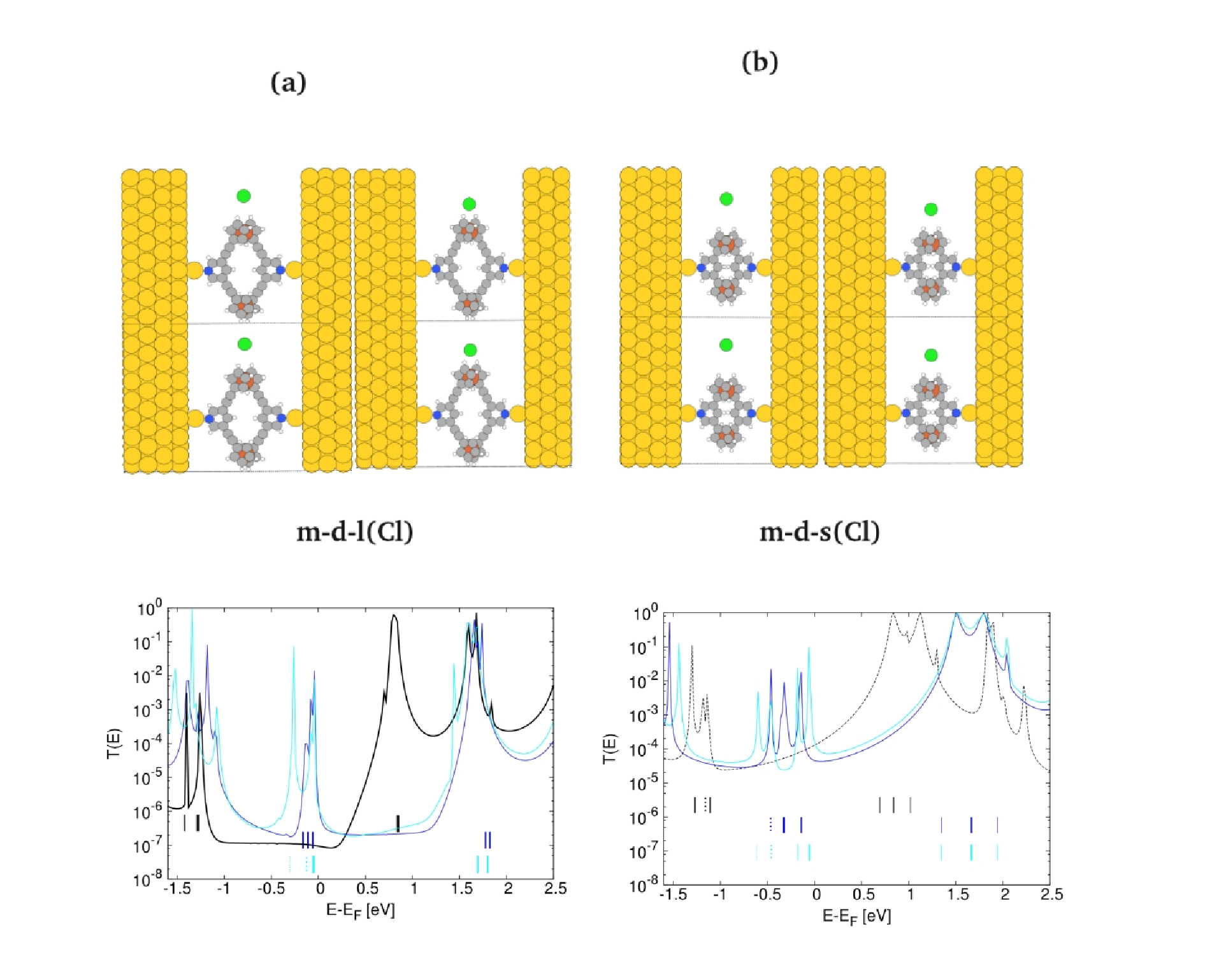}
    \caption{\small Junction geometries for two neighbouring cells in the periodic setup for the scattering region (upper panels) and T(E) from NEGF-DFT calculations (lower panels) for the branched molecules a) $m-d-l$ and b) $m-d-s$ where the distance d$_{Cl-Fe}$ between the Fe atom on one branch and the chloride counter ion stabilizing the positive charge on the respective junction has been varied. For the transmission functions in the lower panels, which we calculated from NEGF-DFT, the neutral reference systems in the absence of charging or chlorine have been marked as in Fig.~\ref{fig3}, i.e. solid black for $m-d-l$ and dashed black for $m-d-s$, while the cyan and blue curves mark T(E) for the charged junctions for an asymmetric (right top panels) and symmetric (left top panels) placement of the chloride ions between the two branches in neighbouring cells, respectively. The eigenenergies of the relevant MOs are also indicated in the lower panels, where the color code reflects the one used for the transmission functions and the line type distinguishes between the two branches.}
    \label{fig10}
    \end{center}
\end{figure*}

\subsection{Methodology for the charging of the molecule in the junction}

In this section we address the effect of the selective charging of the ferrocene center on one of the two branches in the two double-branched molecules m-d-l and m-d-s on the conductance in order to assess their usefulness as molecular switches along the lines suggested in the introduction. While in experiments one of the two ferrocene moieties has to be marked by a substituent in order to achieve the asymmetry allowing for redox splitting~\cite{tim}, in our theoretical calculations we can achieve the same effect by making use of an idea introduced in Ref.~\cite{lambert4} where the electronic structure of a benzene molecule was distorted in an asymmetric fashion by the strategic placement of a potassium point charge. In our work we use a method for the charging of the branched compounds with a chlorine atom in the cell close to the molecule which due to its higher electronegativity absorbs an electron from the junction while oxidizing it in the process~\cite{pyridil4} while the overall neutrality of the device region is still maintained. As we describe in detail in Ref.~\cite{pyridil4} where we introduced this approach for the oxidation of another organometallic complex, the generalized $\Delta$ self-consistent field ($\Delta$ SCF) technique~\cite{dscf1,dscf2} has to be applied in such a setup for ensuring that the self interaction problem of DFT is defied and the chloride ion is charged with one full electron while the resulting positive counter charge is distributed across the molecule and surfaces of the leads.

\begin {table}
\setlength{\tabcolsep}{0.5em}
\caption {Partial charges in units of fractions of 1 e as obtained from a Bader analysis~\cite{bader} for the neutral and charged junctions defined in Fig.~\ref{fig10}, where Fc 1 and Fc 2 denote the ferrocene closer to and further away from the chloride ion, respectively. The conductance G for all junctions as defined by T(E$_F$) in Fig.~\ref{fig10} is given in units of G$_{0}$.} \label{tab:charge}
\begin{center}
    \begin{tabular}{|c|c|c|c|}
    \hline
     & Fc 1 & Fc 2 & G \\ \hline
    \hline
    $m-d-l$ (neutral)  & -0.41 & -0.44 & 0.95\texttimes 10$^{-7}$ \\
    \hline
    $m-d-l$ (d$_{Cl-Fe}$=5.7\AA)  & -0.61 &  -0.64 & 1.89\texttimes 10$^{-6}$ \\
    \hline
    $m-d-l$ (d$_{Cl-Fe}$=4.3\AA)  & -0.71 &  -0.55 & 1.41\texttimes 10$^{-6}$ \\
    \hline
    \hline
    $m-d-s$ (neutral)  & -0.17 & -0.17 & 1.28\texttimes 10$^{-4}$ \\
    \hline
    $m-d-s$ (d$_{Cl-Fe}$=7.2\AA)  & -0.28 & -0.23 & 4.50\texttimes 10$^{-5}$ \\
    \hline
    $m-d-s$ (d$_{Cl-Fe}$=5.4\AA)  & -0.36 & -0.18 & 1.57 \texttimes 10$^{-4}$ \\
    \hline
    \end{tabular}
\end{center}
\end {table}

Following the concepts of Ref.~\cite{lambert4} we built unit cells for the device region with a 4 $\times$ 8 over-structure in the surface plane in order to create some space to vary the position of the chloride ion in one direction but with the reduction of the unit cell length along the other lattice vector keep the computational costs on a reasonable level. Since the position of the chloride anion in the unit cell has a marked influence on the distribution of the positive charge on the molecule and surface due to electrostatic attraction~\cite{pyridil4}, we vary the distance of the ion to one of the two ferrocene centers as d$_{Cl-Fe}$ (Fig.~\ref{fig10}) in order to create asymmetry and denote the closer one as Fc 1 and the one further away as Fc 2 in the following. Because of the different sizes of molecules $m-d-l$ and $m-d-s$, the detailed values of d$_{Cl-Fe}$ also differ for the two cases, with values of 5.7\AA{} and 7.2\AA{} for the symmetric setup where the ion has an equal distance to both Fe atoms and of 4.3\AA{} and 5.4\AA{} where it is markedly closer to that of Fc 1.

\subsection{Partial charge distributions}

In Table~\ref{tab:charge} we list the resulting partial charges on Fc 1 and Fc 2 as obtained from a Bader analysis~\cite{bader} where it can be seen that already in the neutral cases without the presence of the chlorine the molecules have some positive partial charges since they lose fractions of electrons to the anchor groups and the gold surfaces. When the chloride ion is introduced into the cell and a negative partial charge corresponding to one electron is enforced on it, only fractions of the resulting positive counter charge reside on the ferrocene moieties, while the partial charge on the surface changes from negative to positive (not shown here), an effect which has been discussed in terms of the respective electronegativities for another metal-organic complex in Ref.~\cite{pyridil4}. We find also a substantial accumulation of negative partial charges on the acetylenic spacers which explains why both charges on the ferrocene groups of molecule $m-d-l$ are consistently more than twice as large as those found for $m-d-s$ with and without charging via the chlorine atom.

For both compounds, however, the partial charge is distinctly higher on Fc 1 than on Fc 2 in the asymmetric setup which is also reflected by the differences in peak shifts in the respective transmission functions in Fig.~\ref{fig10}. While the peaks in the LUMO region are almost rigidly shifted to higher energies as a consequence of the charging for both molecules regardless of whether the ion is placed symmetrically or asymmetrically with respect to the Fe positions, there are distinct differences in the HOMO region where the asymmetry induces peak splitting which could be expected from the discussion in Sec.~\ref{sec:dft} where we noted that the HOMOs are mostly localized on the ferrocene moieties and the LUMOs on the pyridyl anchors. 

\subsection{Transmission functions and DQI for the charged compounds}

Our expectation from T(E) for the neutral molecules in Sec.~\ref{sec:dft} was that due to the flat behavior of the function in the HOMO-LUMO gap induced by the narrowness of the HOMO peak and DQI close to the LUMO peak, there would be almost no change in the conductance as a consequence of charging for system m-d-l, while the Lorentzian decay of the LUMO peak for m-d-s might give rise to charge induced conductance changes since the Fermi level would move down the tail of the peak. These assumptions assumed a rigid shift of T(E) and did not foresee that the HOMO-LUMO gap is reduced in size by the charging where the tails of the HOMO peak now play a more active role for the definition of the conductance as can be seen from the NEGF-DFT calculations for the charged systems in Fig.~\ref{fig10} where we also list the corresponding values for G in Table~\ref{tab:charge}. 

It can be seen that for m-d-s the transmission functions of the neutral and the asymmetrically charged system cross each other almost exactly at E$_F$ resulting in almost equal conductance values, while the conductance is enhanced by the charging for m-d-l where the Fermi level is now at the shoulder of the HOMO peak for both the symmetric and the asymmetric setups. This latter charging effect on the conductance for m-d-l, however, would only result in an ON/OFF ratio of $\sim$ 15-20 which is by far too small for an operative transistor. Moreover, our initial idea that the charging might have an influence on the presence or absence of DQI effects is not supported by the changes in the transmission function, although the DQI induced flattening of the LUMO peak seems to be somewhat reduced for m-d-l in the cyan curve in Fig.~\ref{fig10} for the asymmetrically charged setup where there is also a corresponding energy splitting found for the LUMO and LUMO+1 which are almost degenerate in the neutral system.

\section{Summary}\label{sec:summary}

In this study we investigated the potential use of branched molecules containing ferrocene centers in two branches as molecular transistors where the switching would be achieved by a redox process allowing to alternate between an ON and an OFF state and the latter might have a substantially reduced conductance due to DQI. We found such a DQI effect in the electron transmission for one of the branched molecules we studied in its neutral state but this effect was not altered significantly enough by charging for enabling a transistor functionality with this particular system. Quite surprisingly, the appearance of the effect was closely linked to the presence of acetylenic spacers between the ferrocene moieties and the pyridyl anchor groups. In an analysis, where we mapped the essential orbital characteristics of the metal-organic compounds under investigation onto more and more simplified tight binding models in a systematic way, we could identify the structural sources for this unexpected finding. The key quantity turned out to be the direct through-space coupling between the anchor groups, which is determined in its size and sign by the detailed three-dimensional conformation of the respective molecule. This is fundamentally different from DQI as described for planar $\pi$ conjugated hydrocarbons, where simple topological rules could be derived recently and where geometrical details of the molecular structure beyond next-neighbor connectivity do not play an essential role. The systematics of our analysis in this work can be applied to other metal-organic compounds exhibiting DQI effects with an influence on their conductance and therefore provides an enabling tool for the rational design of molecular transistors. 

\begin{acknowledgments}
All authors have been supported by the Austrian Science Fund FWF (project number No. P27272). We are indebted to the Vienna Scientific Cluster VSC, whose computing facilities were used to perform all calculations presented in this paper (project No. 70671). We gratefully acknowledge helpful discussions with Tim Albrecht and Michael Inkpen.  
\end{acknowledgments}

\bibliographystyle{apsrev}

\bibliographystyle{apsrev}

\end{document}